# An Acoustic Demonstration of Galileo's Law of Falling Bodies

Michael W. Courtney[i] and Elya R. Courtney[ii]

[i]U.S. Air Force Academy, 2354 Fairchild Drive, USAF Academy, CO, 80840-6210
Michael.Courtney@usafa.edu

[ii]BTG Research, PO Box 62541, Colorado Springs, CO 80962-2541

**Introduction**

An acoustic method is presented for analyzing the time of falling motion.  A ball is dropped from a measured height.  The dropping device makes a distinct sound a well-determined time (roughly 14 milliseconds) after release.  The ball subsequently makes a second distinct sound when it hits the surface below.  These sounds are captured with a microphone resting on the surface and are readily apparent in the acoustic waveform.  At each height (0.25m, 0.50m, 0.75m, and 1.00m), the measured drop time agrees with the drop time predicted by the law of falling bodies with a typical accuracy of 4.3 ms.

Since Galileo, analysis of falling motion has been challenged to measure time with suitable accuracy to demonstrate the Law of Falling Bodies.[1]  Galileo employed ramps to slow descent of bodies; Atwood invented the Atwood machine.[2]  Various electronic techniques have been implemented to measure accelerations, including the 60-Hz spark technique,[3] photogates,[4] and high-speed video.  The accuracy of these methods can be impressive, but requires specialized equipment.  Furthermore, these techniques are often unable to determine the instant of drop with sufficient accuracy, thus limiting the analysis to expectations of linear increase of velocity vs. time rather than the quadratic increase of the distance vs. time.  The acoustic method presented here uses commonly available equipment and offers a more direct test of the law of falling bodies; namely, that the distance fallen is proportional to the square of the time and the acceleration of gravity.

**Method**

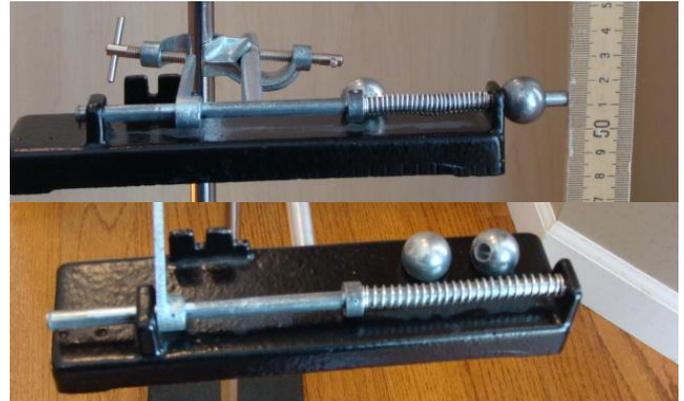

Figure 1: Ball drop apparatus showing ball held in place by rod (top), and lever in position striking stop which makes distinct sound right after ball is released (bottom).

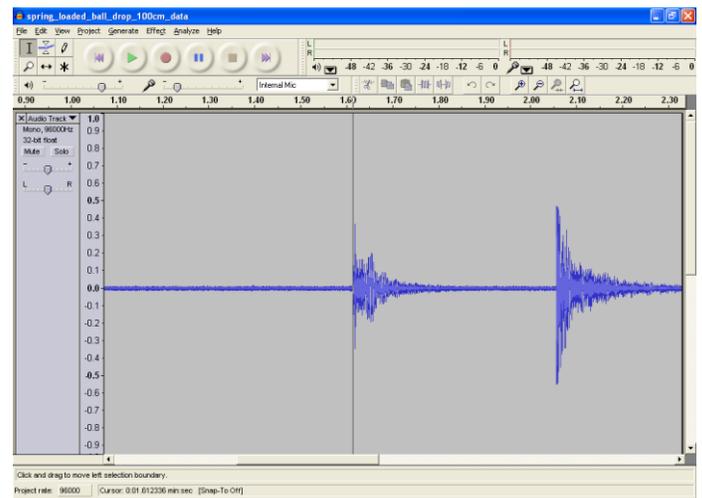

Figure 2: Audacity screen showing distinct sounds of lever striking stop (first peak) and ball striking floor (second peak).

A steel ball of diameter 1.8 cm is supported by a horizontal rod that can be retracted by raising a lever on a spring loaded device as shown in Figure 1.  The height of the ball above the surface (table or floor) is measured with a meter stick.  When the lever is lifted, the spring loaded device withdraws the rod supporting the ball, allowing it to fall.  A short time after the ball begins to fall, the lever strikes a metal stop on the device, making a distinct sound.

A microphone is set on the surface near where the ball is expected to land. The microphone picks up the distinct sound of the lever striking the metal stop right after the ball is released and the sound of the ball striking the landing surface. The digitized sound is recorded with the Audacity[5] software program. A typical waveform is shown in Figure 2. The fall time can be estimated from the time difference between these two distinct sounds. Five waveforms were recorded and five times were determined for each height. The average times and standard deviations from the mean were then computed for each height. (This method ignores the delays of each sound reaching the microphone. These delays are less than 4 ms and can be mitigated by placing the microphone equidistant from the drop height and striking surface.)[6]

**Data and Analysis**

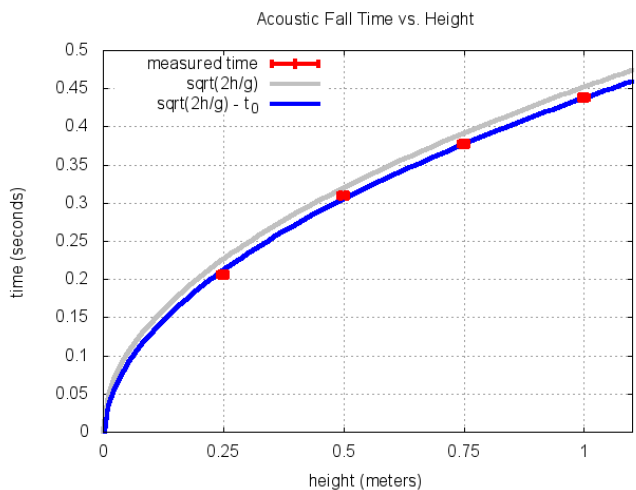

Figure 3: The grey line is the theoretical fall time in the absence of air resistance. The blue line is a best fit line which subtracts the delay between ball drop and the noise of the lever hitting the metal stop on the apparatus.

The measured times and theoretical models are shown in Figure 3. The vertical error bars represent the standard deviation from the mean and the horizontal error bars represent a generous estimate of the height uncertainty of 1 cm. A red line (top) shows the theoretical fall time vs. height. The blue line (bottom) subtracts a constant time from the red line to account for the delay between ball drop and the sound of the lever hitting the metal stop. This delay is determined by fitting the curve to the delay ($t_0$) which gives a delay of 14.2 ms and a root mean square of residual errors of 4.3 ms.

**Conclusion and Discussion**

An acoustic method for demonstrating Galileo's Law of Falling Bodies fits the expected predictive model with a typical drop time accuracy of 4.3 ms, which corresponds to 1.4% of the fall time from 0.5m. This technique directly determines fall time for a given distance and uses much less specialized equipment than other techniques commonly employed in introductory physics laboratories.

It is also possible to adapt this acoustic technique by simply rolling a steel ball or glass marble off of a horizontal surface and determining the time delay between cessation of the rolling sound and the sound of the object striking the surface below. The challenge in this alternative method is that most surfaces continue to resonate for a short time after the sphere leaves. Skillful dampening of the sound on the rolling surface can minimize this and a comparable analysis technique regarding the delay in cessation of sound yields comparable results with those shown here. A more accurate technique would be to use a metal ball to complete a circuit which would be opened the exact instant the ball is dropped. However, this significantly increases the methodological complexity and requires synchronization between the ball drop timing and the acoustic determination of the instant the ball hits the surface below.